\documentclass[letter]{aa} 

\newcommand{\XMM}{\textit{XMM-Newton}}    
\newcommand{\CXO}{\textit{Chandra}}     

\usepackage{graphicx}
\usepackage{txfonts}
\begin{document}

   \title{X-ray emission from an FU Ori star in early outburst: HBC 722}


   \author{Armin Liebhart
          \inst{1}
          \and
          Manuel G\"udel
	  \inst{1}
          \and
          Stephen L. Skinner
	  \inst{2}
          \and
          Joel Green
	  \inst{3}
          }

   \institute{Department of Astrophysics, University of Vienna,
       T\"{u}rkenschanzstrasse 17, A-1180 Vienna, Austria\\
              \email{armin.liebhart@univie.ac.at; manuel.guedel@univie.ac.at}
         \and
             CASA, University of Colorado, Boulder, CO 80309-0389, USA\\
             \email{stephen.skinner@colorado.edu}
         \and
             Department of Astronomy, The University of Texas at Austin, 2515 Speedway, Stop C1400, Austin, TX
	     78712-1205, USA\\
             \email{joel@astro.as.utexas.edu}
             }

   \date{Received August 10, 2014; Accepted September 17, 2014}

 
  \abstract
   {}
   {We conducted the first X-ray observations of the  newly erupting FU~Ori-type outburst in HBC~722 (V2493 
   Cyg) with the aim to characterize its X-ray behavior and near-stellar environment during early outburst.}    
   {We used data from the \XMM\ and \CXO\ X-ray observatories to measure X-ray source temperatures and luminosities as
   well as the gas column densities along the line of sight toward the source.}
   {We report a \CXO\ X-ray detection of HBC~722 with an X-ray luminosity of
   $L_{\rm X}\approx 4\times 10^{30}~\mathrm{ergs~s}^{-1}$. 
   The gas column density  exceeds values expected from optical extinction and standard gas-to-dust ratios.
   We conclude that dust-free gas masses are present around the 
   star, such as strong winds launched from the inner disk, or massive accretion columns.
   A tentative detection obtained by \XMM\ two years earlier after an initial optical peak revealed a fainter
   X-ray source with only weak absorption.}
   {}

   \keywords{stars: individual: HBC~722 -- stars: pre-main sequence -- X-rays: stars
               }

   \maketitle
%
\section{Introduction}

The small group of FU Ori stars (FUors) are pre-main sequence stars showing giant, 
long-lasting optical outbursts \citep{1977ApJ...217..693H}. 
The defining photometric feature of an FUor  
is a dramatic increase in optical brightness $(\Delta V\gtrsim 4~\mathrm{mag})$, 
followed by a slow decline that can last from decades to centuries. 
This evolution is accompanied by a change from a typical T Tauri stellar spectrum
to an F--G supergiant spectrum in the optical and to a late-type (K and later) giant spectrum 
in the near-infrared. Other important features are heavily
blueshifted absorption lines (indicating velocities $>100$~km~s$^{-1}$), CO
overtone absorption bands around 2~$\mu$m, and almost no lines in emission,
with the significant exception of H$\alpha$. The few emission lines show P Cygni profiles, 
which indicates strong winds, again especially in $\mathrm{H\alpha}$. Furthermore, FU Ori 
stars in outburst are  accompanied by newly formed reflection nebulae. 
The theory for FUor outbursts assumes a cataclysmic accretion event in which the
accretion rate increases by a factor of 100 or more during a relatively short time interval 
$(\lesssim100\,\mathrm{years};$ e.g., \citealt{1994ApJ...427..987B, 2006ApJ...650..956V}). 

A second group of eruptive variables,  EXors (after their prototype EX Lup, \citealt{2008AJ....135..637H}) 
exhibit similar observational features albeit with more modest amplitudes and on shorter time 
scales of several months to a few years. Optically, an EXor appears 
as a T Tauri star, while a FUor shows a much broader optical and near-IR peak that is dominated by 
the hot inner disk that outshines the central star.

X-ray studies of FUors and EXors are interesting because strong accretion may change 
the magnetic topology in the stellar  environment. 
The EXor V1647 Ori revealed a rapid increase in the X-ray flux 
by a factor of $\approx 30$ that closely tracked the optical and near-infrared light curves 
\citep{2006ApJ...648L..43K}. The X-ray spectra hardened during the peak and softened 
during the decay. Extremely high temperatures of about $6\times 10^7$~K indicated magnetic
reconnection in star-disk magnetic fields.
In contrast, V1118 Ori showed X-ray variations by no more than a factor of two during 
its 1--2 magnitude optical brightness increase on time scales of 50~days, but instead
revealed a remarkable softening during the outburst, which indicates that the hot plasma  
disappeared during that episode \citep{2005ApJ...635L..81A}.

X-rays from FUors have been more elusive. Although two classical FUors have been recorded 
in X-rays so far, namely the prototype FU Ori \citep{2006ApJ...643..995S} and V1735 Cyg
\citep{2009ApJ...696..766S}, these observations occurred long after the initial outbursts,
during the gradually declining phase. Both revealed a high-temperature plasma ($kT > 5$~keV),
which has been assumed to be the result of coronal activity. The X-ray properties  resemble   
those of Class I protostars.
Two other FUors, V1057 Cyg and V1515 Cyg, remain X-ray nondetections \citep{2009ApJ...696..766S}.  

Here, we report the first X-ray detection of an FUor in the initial stages of 
its outburst. The recently erupting  HBC~722 is the first and so far only 
FUor that has been monitored from its early outburst phase 
to the (presumably) main peak in all available wavelength bands 
\citep{2012ApJ...755..157D,2013JKAS...46..253S,2013ApJ...764...22G,2014BlgAJ..20...59S}.
We obtained three X-ray observations during the initial rise of the optical light, 
during a subsequent minimum, and during the following second maximum.
HBC 722 has been frequently studied in the optical and 
near-infrared and has meanwhile been classified as a genuine FUor \citep{2012A&A...542A..43S}.
The interest in its high-energy radiation was initially spurred by its
apparent detection with  {\it SWIFT}  \citep{2010ATel.3040....1P}, 
although we caution that - as demonstrated below - the field around our target 
is very complex and crowded in X-rays for the \XMM\ and Swift angular resolution. 
There are no pre-outburst X-ray observations of HBC~722.

\section{HBC~722}

The newly detected FUor HBC~722 (V2493 Cyg, LkH$\alpha$ 188 G4) \citep{2010A&A...523L...3S} is located    
at a distance of 520~pc \citep{2013ApJ...764...22G}
and is only the second FUor that has been observed prior to its outburst. 
Its pre-outburst characteristics \citep{1979ApJS...41..743C} indicate that it was an emission-line star 
in the spectral range K7-M0, that is most likely a classical  Tauri star (CTTS).

\begin{figure}[t!]%
\begin{center}
\includegraphics[scale=0.38, angle=-90]{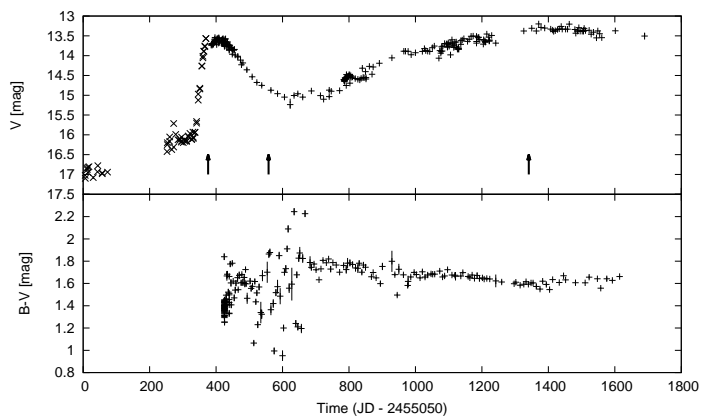}
\caption{$V$ (top) and $B-V$ (bottom) light curves of HBC~722 from {\tt aavso.org} ($+$ symbols), complemented 
by $R$-band data from \citet{2011ApJ...730...80M} (top, early rise phase, $\times$ symbols). The three arrows mark
the  XMM1, XMM2, and CXO1 observing dates (left to right; Table~\ref{tab:obs}).\label{fig:b-v}}
\end{center}
\vskip -0.5truecm
\end{figure}

HBC~722 has become an object of great interest because its eruption was detected during its initial phase,
and excellent photometric and spectroscopic pre-outburst data are available (Sect.~\ref{sect:res}).
Its luminosity started to increase in May 2010 and reached a first
maximum in October 2010 (Fig.~\ref{fig:b-v}), making it the fastest rise ever recorded for an FUor 
\citep{2012A&A...542A..43S}. After a very fast initial decline until April
2011,  it started to increase again and only recently reached a plateau. HBC~722 exhibits
all defining features of a classical FUor. Its bolometric luminosity, $L_{\rm bol}$,
increased from $0.7\,\mathrm{L_{_{\odot}}}$ to $12\,\mathrm{L_{_{\odot}}}$ 
\citep{2011ApJ...730...80M}, which is an increase at the lower end for the class. 
The calculated accretion rate of HBC~722
of $\sim 10^{-6} M_{\odot}$~yr$^{-1}$ also lies at the lower end of the class
\citep{2011A&A...527A.133K}. But like other FUors, HBC~722 changed its spectrum 
from typical CTTS characteristics to a G3 supergiant in the optical and a K-type giant
in the near- to mid-infrared \citep{2012A&A...542A..43S}. 
It only shows one detected emission line, namely $\mathrm{H}\alpha$ 
\citep{2011BlgAJ..17...88S}. The accompanying reflection nebula has grown to $\sim 2400~\mathrm{AU}$ in 2011 \citep{2011ApJ...730...80M}.

\section{Observations, data reduction, and analysis}

We requested \XMM\ \citep{2001A&A...365L...1J} target of opportunity time twice to obtain an early  
X-ray view of HBC~722 (observations XMM1/2). 
Because of the complex source region with its many faint X-ray  sources, we additionally obtained    
\CXO\ X-ray Observatory \citep{2000SPIE.4012....2W} guest observer time 
(observation CXO1).
Table~\ref{tab:obs} summarizes dates, exposure times, 
number of counts in the source extraction area and net counts
after background subtraction, and 
detection status of all observations; the X-ray observation times are marked in Fig.~\ref{fig:b-v}.

For the \CXO\ observation we used the {\it Advanced CCD Imaging
Spectrometer} ({\it ACIS-S} \citep{2003SPIE.4851...28G} in {\tt vfaint} mode. We reprocessed the level 2 event 
file using \CXO\ Interactive Analysis of Observations (CIAO) vers. 4.4, applying calibration data from 
CALDB version 4.4.8. Source detection was performed with the CIAO task {\tt wavdetect}.
We extracted the 0.5-10~keV source spectrum of the clearly detected HBC~722 using a circular 
source area with radius = 1.76$^{\prime\prime}$, and a background spectrum from a large
source-free area on the same CCD, using the CIAO {\tt specextract}
tool that also delivers the response matrix. 

The \XMM\ observations were obtained by the {\it European Photon Imaging Cameras (EPIC).}
Given the faintness of the possible HBC~722 source, we report
only 0.3-7~keV results from the pn camera \citep{2001A&A...365L..18S}, but the MOS cameras 
also show a marginal excess above background at the HBC~722 position. Data were reprocessed with the  
{\it Scientific Analysis System} [SAS, version 12.0.1\footnote{See ``User Guide to the \XMM\ Science Analysis System'', 
Issue 7.0, 2010 (ESA: \XMM\ SOC)}]. We performed source detection using the CIAO 
{\tt wavdetect} task with a point spread function with a 40\% encircled energy radius 
of 6$^{\prime\prime}$\footnote{see XMM-Newton Users Handbook Issue 2.1, Sect. 3.2.1.1.}.
We extracted spectra for the source (within 
 4.5$^{\prime\prime}$ around the best-matching XMM2 source) and the background (from 
a large, source-free area) using the task {\tt evselect}.
The programs {\tt rmfgen} and {\tt arfgen} created the redistribution matrix (rmf) and ancillary 
response files (arf).
\begin{table}[t]
\setlength{\tabcolsep}{2pt}
\caption{Log for HBC~722 X-ray observations}
\begin{center}\label{tab:obs}
\begin{tabular}{p{1.2cm}p{1.7cm}p{1.7cm}p{1.5cm}p{1.3cm}p{0.6cm}}
\hline 
\hline 
Obs.        & ObsID      & Date       &  Exp.           & Source      & Det.  \\
code        &            & (y/m/d)    &  time$^a$ (ks)  & counts$^b$  &         \\
\hline 
XMM1        &  656780701 & 2010/11/25 &  16.85      &  -           & no\\
XMM2        &  656781201 & 2011/05/26 &  17.56      &  19.3$\pm 6.1$ & yes?\\
CXO1        &  14545     & 2013/07/17 &  29.68      &  19.1$\pm 4.6$ & yes\\
\hline 
\end{tabular}
\tablefoot{
$^a$ Exp. time corresponds to LIVETIME;
$^b$ net source counts in $0.3-7$~keV band for XMM2 (pn camera); $0.5-10$~keV for CXO1.
}
\end{center}
\end{table}

\begin{figure}[t]%
\includegraphics[scale=0.22]{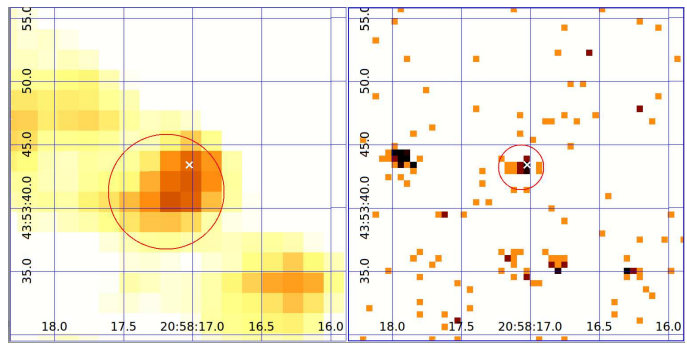}
\caption{Maps for XMM2 (left; pixel size 1.6$^{\prime\prime}$, energy range = 0.3-7~keV) and CXO1 observations 
(right; pixel size 0.49$^{\prime\prime}$, energy range = 0.5-10~keV). The circles show the extraction regions 
(radii = 4.5$^{\prime\prime}$ and 1.76$^{\prime\prime}$ for XMM2 and CXO1, respectively), 
centered at the {\tt wavdetect} source coordinates. The crosses mark 
the 2MASS position \citep{2003yCat.2246....0C}. The \XMM\ image has been smoothed with a 
Gaussian.\label{fig:xraypictures}}
\vskip -0.2truecm
\end{figure}%

We modeled the observed spectra with {\tt XSPEC} \citep{1996ASPC..101...17A}, using
a combination of a gas absorption column density ({\tt wabs} model) and a spectrum of a
collisionally ionized plasma ({\tt vapec} model). The element abundances were adopted from the XEST project
\citep{2007A&A...468..353G}, corresponding to typical values for pre-main sequence stars.
Our model thus delivers the gas column density $N_{\rm H}$ along the line of sight toward
the emitting source, the (average) source temperature, $T$,
and a volume emission measure, EM. We derived the flux by integrating  over the
energy range 0.3-10 keV and calculated the X-ray luminosity $L_{\rm X}$ using a stellar 
distance of $d = 520$~pc.
We conducted a 2 D 
parameter study by evaluating the best fits for any given combination of $N_{\rm H}$ and $kT$, and by determining
the 90\% confidence level for the two parameters of interest.

\section{Results}\label{sect:res}

The CIAO task {\tt wavdetect} found a 7.9$\sigma$ X-ray source   
in CXO1 comprising 20 counts at 
RA(2000.0) = 20h 58m 17.07s$\pm 0.01$s, 
$\delta$(2000.0) = 43$^{\circ}$ 53$^{\prime}$ 43.21$^{\prime\prime}\pm 0.05^{\prime\prime}$,
offset by only $\approx 0.50^{\prime\prime}$
from the expected position, 
RA(2000.0) = 20h 58m 17.025s$\pm 0.006$s, 
$\delta$(2000.0) = 43$^{\circ}$ 53$^{\prime}$ 43.39$^{\prime\prime}\pm 0.03^{\prime\prime}$ 
\citep{2003yCat.2246....0C}\footnote{The rms positional uncertainty for an 
on-axis point source is $\approx 0.42^{\prime\prime}$; see CXO Users Manual, http://asc.harvard.edu/proposer/POG.}.
The \XMM\ observation XMM1 did not reveal any significant source at the
expected position of HBC~722, but the region is, at the angular resolution of \XMM,
very crowded and potentially blurred by other point-like X-ray sources (Fig.~\ref{fig:xraypictures}). 
{\tt Wavdetect} revealed a point-like 3$\sigma$ X-ray source in XMM2 at 
RA(2000.0) =20h 58m 17.19s$\pm 0.038$s,
$\delta$(2000.0) = 43$^{\circ}$ 53$^{\prime}$ 41.31$^{\prime\prime}\pm 0.36^{\prime\prime}$, offset from
the expected position by  2.7$^{\prime\prime}$, corresponding to $\approx 0.7\sigma$ rms uncertainty of 
the absolute \XMM-pointing accuracy\footnote{The rms positional uncertainty for an on-axis point source is $\approx 4^{\prime\prime}$;  
see \XMM\ Users Handbook, http://xmm.esac.esa.int/external/xmm$\_$user$\_$support/documentation/
uhb/index.html.}. 
Similar offsets were found for bright sources in the field.
Because of some potential contamination by neighboring sources, we consider this faint detection tentative but useful 
in the context of the later \CXO\ detection.

\begin{figure}[t]
\includegraphics[width=0.50\textwidth]{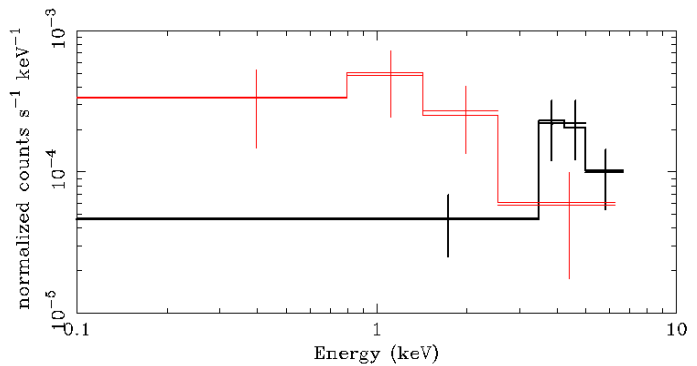}
\caption{Observed spectra (error bars) and fits (histograms) of CXO1 (black) and XMM2 (red), 
binned to a minimum of 5 and 7 cts/bin, respectively.     
\label{fig:chandraspec_1}}
\end{figure}

Table~\ref{tab:xspec} shows the results for the XMM2 and CXO1  observations;
$L_{\rm X}$ is the unabsorbed luminosity in the 0.3--10~keV band. 
The {\tt XSPEC} norm is defined as $\mathrm{EM}/(4\pi 10^{14}d^2)$, $d$ being the distance to the star.
The near-absence in the \CXO\ spectrum of X-ray counts below $\approx 3$~keV and a spectral 
peak around 4--5~keV (Fig.~\ref{fig:chandraspec_1}) require very high $N_{\rm H}$. 
Our best fit to a spectrum rebinned to at least five counts per bin delivered    
$N_{\rm H} \approx 1.4\times 10^{23}$~cm$^{-2}$ (90\% confidence range:
$(4.4-56)\times 10^{22}~\mathrm{cm}^{-2}$).
For a standard interstellar gas-to-dust ratio, we expect an optical extinction, $A_{V}$, of approximately
$A_{V} \approx N_{\rm H} / (1.8\times 10^{21}~{\rm cm}^{-2}) \approx 80$~mag or higher, in disagreement 
with optically determined $A_{\rm V}$ measurements, which are sensitive primarily
to dust absorption (see  below). Given the few counts
for the $\chi^2$ statistics, we alternatively used unbinned data in conjunction with the C statistic
(\citealt{1979ApJ...228..939C}, Table~\ref{tab:xspec}). We fully confirm the high $N_{\rm H}$ and found very 
similar 90\% confidence ranges for all parameters as for the binned data.

\begin{figure}[t]
\includegraphics[width=0.24\textwidth,angle=-90]{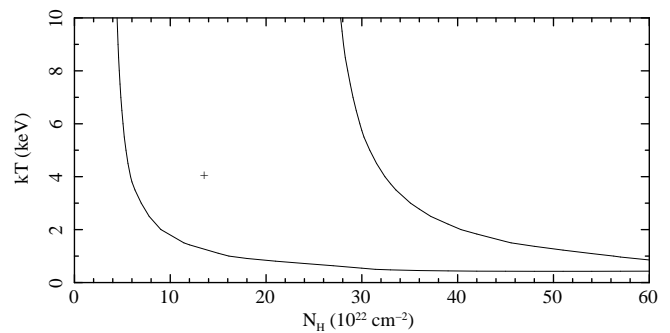}
\caption{90\% confidence level range for $kT$ vs $N_{\rm H}$ for the CXO1 observation using binned
data. The cross marks the best fit. \label{fig:contour}}
\end{figure}

\begin{table}[t!]
\setlength{\tabcolsep}{2pt}
\caption{Results for the {\tt XSPEC} spectral model fits}
\begin{center}\label{tab:xspec}
\begin{tabular}{p{2.35cm}p{1.9cm}p{2.15cm}p{2.0cm}}
\hline 
\hline 
Parameter & XMM2 (pn) & CXO1 (ACIS) & CXO1 unbin$^a$ \\
\hline 
$N_{\rm H}/10^{22}~\mathrm{cm^{-2}}$                  & $0.08(...,0.72)$ 
                                                      & $13.8(4.4,55.9)$  
						      & $20.5(3.3,55.2)$  \\
$kT ~\mathrm{[keV]}$                                  & $8.4(1.3,...)$
                                                      & $4.1(0.55,...)$ 
						      & $2.3(0.55,...)$\\
norm$/10^{-5}~\mathrm{cm^{-3}}$                       & $0.9(0.4,2.3)$   
					              & $7.3(1.6,1840)$
						      & $20.4(1.6,1900)$  \\ 
$L_{\rm X, best}~\mathrm{[erg~s^{-1}]}$               & $4.3\times 10^{29}$
                                                      & $3.5\times 10^{30}$  
						      & $7.8\times 10^{30}$\\
Statistic/dof                                         & $0.04/1$     
                                                      & $0.04/1$
						      & $17.9/17$ \\
\hline
\end{tabular}
\tablefoot{Norm = EM/$(4\pi 10^{14} d^2)$, where $d$ = distance to the star (520 pc) 
and $L_{\rm X, best}$ is the unabsorbed $L_{\rm X}$ for the best fit.
Statistic/dof is  $\chi^2$ for Cols. 2 and 3, and $C$ for Col. 4, divided by the number of degrees of freedom; 
90\% confidence ranges (for one parameter of interest) are given in parentheses. 
Ellipsis dots indicate an unconstrained error range.  $^a$ For unbinned data.}
\end{center}
\vskip -0.5truecm
\end{table}

To assess the observed $A_{V}$,  we extracted $B-V$ colors from the online 
data of the {\it American Association of Variable Star Observers} (AAVSO; \citealt{henden2014}). 
$B-V \approx 1.7$~mag remains nearly constant during the recording time     
(Fig.~\ref{fig:b-v} bottom). The G3 supergiant optical 
spectrum of HBC 722 \citep{2012A&A...542A..43S} corresponds to $(B-V)_{0} \approx 0.9$~mag 
\citep{1998gaas.book.....B}. We then used
$A_{V}=R_{V} \times E(B-V)$, where $E(B-V)=(B-V)-(B-V)_{0}$ is the color excess and $R_{V}$ is the 
total-to-selective extinction. Applying the standard value for $R_{V}=3.1$ 
\citep{1975A&A....43..133S}, we obtain $A_{V} \approx 2.5$. Using
$E(B-V)=N_{\rm H}/5.8\times 10^{21}~\mathrm{cm}^{-2}$
\citep{1998gaas.book.....B} for a standard interstellar gas-to-dust mass ratio, we 
expect only $N_{\rm H} \approx 4.6 \times 10^{21}$~cm$^{-2}$.

The XMM2 results were derived from a spectrum binned to at least seven counts per bin. 
While the highest source temperature is essentially 
unconfined, the presence of soft photons in the spectral range of 0.5--1~keV requires $N_{\rm H}$ 
to be moderate; we find $N_{\rm H} \approx 8\times 10^{20}$~cm$^{-2}$, with 90\% confidence 
upper limits around $N_{\rm H} \approx 7.2\times 10^{21}$~cm$^{-2}$.  No acceptable joint fit could be found for the
XMM2 and CXO1 observations, demonstrating that the two $N_{\rm H}$ ranges are mutually exclusive.
For standard interstellar gas-to-dust ratios, we expect $A_V$ of about 1~mag 
for the best fit, in agreement with $B-V$ observations and in stark contrast to the CXO1 results.

\section{Discussion and conclusions}
The anomalously high gas absorption (CXO1) in the presence of rather weak optical extinction is similar to an observation in a class of strongly accreting T Tauri stars that exhibit a combination of 
two X-ray spectra, a soft component from a cool ($\approx$2~MK), only weakly absorbed source and a very hard 
component from a hot, strongly absorbed ($N_{\rm H} > 10^{22}$~cm$^{-2}$) source (two-absorber X-ray 
or TAX phenomenology, e.g., \citealt{2008A&A...478..797G}). The classical FUor FU Ori shares these characteristics 
\citep{2006Ap&SS.304..165S, 2010ApJ...722.1654S}. The soft source in DG Tau has been identified with an 
X-ray jet close to the star \citep{2008A&A...478..797G}, while for FU Ori a companion may at least partly explain it \citep{2010ApJ...722.1654S}. In  all cases, however, the hard source, attributed to a magnetically confined 
plasma (e.g., a corona), requires much higher gas column densities than expected from visual extinction and 
a standard interstellar gas-to-dust mass ratio. The proposed models either involve dust-depleted accretion 
streams from the disk to the star or dust-depleted winds launched from the inner disk.

To assess the plausibility of these models, we first estimated $N_{\rm H}$ from accretion streams using mass 
conservation for a stationary flow approximated to be isotropic and radial,
\begin{equation} \label{eq:N_H}
n_{\rm H}=\frac{\dot{M}}{4\pi\ r^{^{2}}\mu m_{\rm p}v}\,,
\end{equation}
where $m_{\rm p} \approx 1.7\times 10^{-24}$~g  and $\mu \approx 1.3$ are the proton mass
and the mean weight per particle for atomic gas.
The mass accretion rate for HBC~722 is $\dot{M}=10^{-6}\ M_{_{\odot}}$~yr$^{-1}$ 
\citep{2011A&A...527A.133K}. 
Halfway between disk border and stellar surface, the accretion stream
velocity will have reached a value of about half the free-fall velocity  
at the stellar surface for a mass element falling from the 
inner-disk border.
Using the inner disk radius  $r=2R_{*}$, $R_* \approx 2 R_{\odot}$, and a stellar mass of 
$M_* = 0.5M_{\odot}$ \citep{2013ApJ...764...22G} we obtain $v \approx 126
$~km~s$^{-1}$ and therefore
$n_{\rm H} \approx 4 \times 10^{12}$~cm$^{-3}$ at $r \approx 1.5R_*$. Integrating over one $R_*$
(from $2R_*$  to the stellar surface) leads to
$N_{\rm H}=5.6 \times 10^{23}$~cm$^{-2}$, within the 90\% confidence range of CXO1 results 
(Table~\ref{tab:xspec}).

We now consider winds launching from the innermost part of 
the disk. Considering the strong heating of the inner disk during an FUor
outburst, we may expect winds, as indeed   
observed in FUors \citep{2003ApJ...595..384H} and in particular also for HBC~722
where they reach velocities of 500 km~s$^{-1}$ \citep{2012A&A...542A..43S}. The gas of   
the inner disk is essentially dust-free because it is far inside the dust
sublimation radius even for a normal CTTS; in any case, the disk 
temperature corresponding to a G supergiant spectrum exceeds the dust 
sublimation temperature. Optical extinction (due to dust) will therefore not be 
enhanced, while X-rays will still be absorbed  by  gas.

If such a dust-poor wind is launched from the inner disk region and expands approximately isotropically, 
then Eq.~\ref{eq:N_H} analogously applies, where $\dot{M}$ now stands for the wind mass-loss rate. 
Integration of $n_{\rm H}$ along the line of sight through the wind from infinity to the disk border 
($2R_*$) provides an estimate for $N_{\rm H}$ if we assume that the wind velocity is constant:
\begin{equation} \label{eq:wind}
N_{\rm H}= \int_{2R_*}^{\infty}\frac{\dot{M}}{4\pi\ r^{^{2}}\mu m_{\rm p}v}dr = \frac{\dot{M}}{8\pi \mu m_{\rm p}vR_*}.
\end{equation}
Using $R_* \approx 2 R_{\odot}$ \citep{2013ApJ...764...22G}, $v \approx 500$~km~s$^{-1}$ \citep{2012A&A...542A..43S}, 
and $\dot{M} \approx 10^{-7}~M_{\odot}$~yr$^{-1}$ (assuming 10\% of the mass accretion rate as for CTTS, 
\citealt{1995ApJ...452..736H}), we find $N_{\rm H} \approx 1.7\times 10^{22}$~cm$^{2}$, somewhat lower than
acceptable for the CXO1 observations, but our model assumptions are fairly crude. 

The XMM1/2 observations do not fit into this picture. At face value, it seems
that strong winds or accretion stream absorption did not prevail in XMM2,
while XMM1 suffered from too much absorption, or the X-ray source was
significantly dimmer. We suggest the following scenario:

The first optical peak was produced by an initial strong disk instability that rapidly led to enhanced accretion and possibly winds that attenuated all  X-rays (XMM1). The initial outburst then ceased, leading into
a more quiet phase during which the star was detected in X-rays (XMM2). 
Subsequently, a gradual increase to a lasting disk instability develops
winds and accretion flows and triggers  enhanced X-ray emission. 
Enhanced X-ray absorption is  now evident (CXO1). 

Direct support for XMM2 picking up a normal CTTS comes from the measured
$L_{\rm X}$. Telleschi et al. (2007) reported statistical correlations
between $L_{\rm X}$ and stellar $L_{\rm bol}$ or mass for a large
sample of CTTS in Taurus. Using $L_{\rm bol} = 0.7L_{\odot}$ and
$M_* = 0.5M_{\odot}$, these best-fit relations (based on two different 
statistical regression methods) all lead to $L_{\rm X} \approx (4.2-5.0)\times
10^{29}$~erg~s$^{-1}$, which agrees well with our XMM2 observation.

Why $L_{\rm X}$ increased by about an order of magnitude during 
the outburst peak is less clear (the increase seems to agree with 
estimates for classical FUors; \citealt{2009ApJ...696..766S, 2010ApJ...722.1654S}). Potential
candidates are reconnection events in dynamo-induced magnetic fields forming 
as a consequence of convection in the strongly heated, unstable disk; the
absorbing medium in this case would be a wind. Alternatively, enhanced
magnetic reconnection in magnetospheric star-disk fields subject to 
increased disturbance by the close-in disk would lead to hard emission, 
while the overlying accretion streams and winds would partially attenuate
the X-rays. Future X-ray monitoring may help to clarify the situation.

\begin{acknowledgements}
We thank an anonymous referee for helpful comments. We acknowledge with thanks the variable star observations from the AAVSO International Database 
contributed by observers worldwide and used in this research. We thank the Project Scientist of \XMM,
Norbert Schartel, for approving our  \XMM\ Target of Opportunity request. 
J.G. and S.S. acknowledge support from Chandra award GO3-14012.
The CXC X-ray Observatory Center is operated by the Smithsonian Astrophysical Observatory for and on behalf of the NASA under 
contract NAS8-03060. This publication is supported by the Austrian Science Fund (FWF).
\end{acknowledgements}

\bibliographystyle{aa}
\bibliography{paperbib}


\end{document}